# 3D Spherical Fluid Antennas for Spatially Reconfigurable Communications

Wenchi Cheng, *Senior Member, IEEE*, Hongyun Jin, *Student Member, IEEE*, and Zhuohui Yao, *Member, IEEE*

***Abstract*—As sixth-generation (6G) wireless systems evolve toward higher frequency bands, large-scale antenna arrays, and intelligent interaction with the wireless environment, conventional fixed-position antennas (FPAs) are increasingly constrained by limited spatial degrees of freedom and insufficient hardware-level adaptability. Fluid antenna systems (FAS) provide new physical-layer flexibility by dynamically reconfiguring antenna ports, geometries, and radiation characteristics. However, existing studies have primarily focused on one- or two-dimensional apertures, leaving the spatial reconfigurability required for complex three-dimensional (3D) propagation environments insufficiently exploited. In this article, we present a 3D spherical fluid antenna system (3D SFAS) architecture for flexible spatially reconfigurable communications. By activating radiating elements in different regions on the spherical surface, the 3D SFAS realizes array-level spatial reconfiguration through flexible switching of reconfigurable regions. Within the selected regions, element-level reconfiguration further adjusts the effective-aperture size, array topology, and radiation characteristics. Through this joint array-level and element-level reconfiguration framework, the 3D SFAS can support highly flexible beamforming, concurrent multi-region transmission, blockage-adaptive aperture switching, effective-aperture reconfiguration, and high-resolution 3D aperture control. We further discuss its potential applications in space-air-ground integrated networks, high-mobility communications, integrated sensing and communication systems, and emergency communications. Finally, numerical results demonstrate the potential benefits of 3D SFAS for improving wireless communication performance through flexible spatial reconfiguration. Overall, the 3D SFAS provides a promising pathway for extending FAS design beyond 2D position switching toward comprehensive 3D spatial reconfigurability.**

## I. Introduction

As sixth-generation (6G) wireless communication systems continue to evolve toward higher-frequency bands, enhanced spatial multiplexing, and intelligent environmental awareness, wireless propagation environments are becoming increasingly dynamic and complex. In particular, emerging scenarios such as space-air-ground integrated networks, holographic communications, integrated sensing and communication (ISAC), and near-field wireless communications demand not only higher data rates and spectral efficiency, but also stronger spatial adaptability to cope with time-varying channels, severe blockage, and signal propagation over multi-dimensional spaces. Conventional massive multiple-input multiple-output (MIMO) and hybrid beamforming technologies have significantly enhanced the spatial signal processing capability of wireless systems. However, the positions, geometric layouts, and electromagnetic characteristics of their antenna arrays are typically fixed after deployment. As a result, traditional fixed antenna arrays, with fixed physical locations and hardware architectures, are gradually approaching their performance limits. They lack spatial repositioning flexibility and therefore struggle to adapt, at the hardware level, to dynamic propagation environments. These limitations have made reconfigurable antennas a key research direction for future wireless systems.

In recent years, fluid antenna systems (FAS) have attracted considerable attention as a promising technology for dynamically reconfiguring the physical properties of antennas [1], [2]. Unlike traditional fixed-position antennas (FPAs), FASs can reconfigure the antenna position, shape, port selection, and radiation pattern, thereby enabling highly flexible reconfigurability [3]. This reconfigurability provides significant advantages under time-varying channel conditions, particularly in wireless communication scenarios involving high-speed mobility, complex propagation environments, and severe blockage. Compared with conventional FPAs, FASs provide a more flexible system architecture that allows for the dynamic selection of radiation regions and the adjustment of radiation characteristics, thus enhancing signal coverage and link quality. In addition, reconfigurable pixel antennas (RPA), as electronically controlled pixel-level reconfigurable antennas, use RF switches to alter the antenna geometry, thereby modifying the current distribution, resonance characteristics, and radiation properties [4]. Compared with traditional fluid antennas, RPAs further advance FAS implementations in terms of switching speed, reconfiguration precision, and radiation characteristics. The integration of pixel antennas into FAS not only provides finer electromagnetic control, but also enables greater adaptability and more precise beamforming in dynamic environments.

However, existing fluid antenna research has primarily focused on port selection over one- or two-dimensional (2D) apertures, without fully exploiting the positional degrees of freedom (DoF) in three-dimensional (3D) space. As a result, current FAS designs do not yet provide the 3D spatial reconfigurability required for 3D propagation and deployment scenarios. Although FASs have improved wireless communication performance through flexible reconfiguration, they still exhibit significant limitations in coping with complex 3D propagation environments expected in future wireless systems, particularly in beam control over 3D angular domains and

This work was supported in part by the National Natural Science Foundation of China (No. 62341132) and in part by QTZX24078.

W. Cheng, H. Jin, and Z. Yao are with the State Key Laboratory of Integrated Services Networks, Xidian University, Xi'an, 710071, China (e-mails: wccheng@xidian.edu.cn, hongyunjin@stu.xidian.edu.cn, and yaozhuohui@xidian.edu.cn).

blockage management under complex propagation conditions. In parallel, recently proposed mechanically movable antenna architectures, such as six-dimensional movable antennas (6D-MA) [5], leverage additional spatial DoFs through physical translation and rotation, thereby providing a new paradigm for spatially reconfigurable communications. However, these methods typically rely on mechanical motion, which introduces issues such as limited response speed, safety and reliability concerns associated with the movement of large-scale base-station equipment, and insufficient positioning accuracy. Therefore, how to further exploit the intrinsic reconfigurability of fluid antennas to realize physical reconfiguration in 3D space remains a key gap in current FAS research.

To address the above limitations, this article proposes a 3D spherical fluid antenna system (SFAS) architecture for future wireless systems. The core idea is to create high-resolution electronically addressable regions on a spherical reconfigurable antenna surface, such that different regions on the spherical surface can be rapidly activated electronically, thereby enabling array-level spatial reconfiguration. Owing to its spherical geometry, the proposed architecture can form reconfigurable effective apertures across both azimuth and elevation dimensions, thereby supporting flexible beamforming, multi-directional access, and blockage-adaptive reconfiguration in 3D environments. Meanwhile, by incorporating element-level reconfiguration in terms of the number of activated radiating elements, effective-aperture size, array topology, and radiation characteristics, the proposed architecture can further achieve fine-grained electromagnetic control. Compared with existing reconfigurable antenna architectures, the proposed architecture offers a promising solution for future spatially reconfigurable wireless systems by enabling faster reconfiguration response, higher reconfiguration precision, and enhanced 3D spatial reconfigurability. Overall, by jointly exploiting array-level and element-level reconfiguration, the 3D SFAS provides a promising pathway for extending FAS design beyond 2D position selection toward comprehensive 3D spatial reconfigurability.

This article aims to provide a comprehensive overview of 3D SFAS from multiple perspectives, including antenna architecture, reconfiguration capability, performance potential, application scenarios, and key challenges. Specifically, we first present the system architecture and basic model of 3D SFAS, and highlight their key advantages over existing 2D FASs and mechanically movable antennas. We then discuss the joint array-level and element-level reconfiguration mechanisms. Furthermore, we analyze their potential performance gains in terms of flexible beamforming, multi-region multi-beam transmission, blockage-adaptive reconfiguration, flexible effective aperture reconfiguration, and high-resolution 3D spatial reconfiguration. Building on this analysis, we further discuss representative application scenarios, including space-air-ground integrated networks, vehicular and high-mobility communications, and ISAC systems. We also summarize open issues related to control optimization, hardware implementation, and system integration. Finally, numerical simulations are provided to demonstrate the potential value of 3D SFAS for future spatially reconfigurable wireless communication systems.

## II. Spherical Fluid Antenna Model and Reconfiguration Architecture

### A. Geometric Model of the Spherical Fluid Antenna

Figure 1 illustrates the system model of the proposed 3D spherical fluid antenna. The basic idea of a 3D SFAS is to extend the reconfiguration principle of conventional 2D fluid antennas from a planar aperture to a spherical surface, thereby exploiting the positional DoFs available in 3D space. Specifically, we consider a reconfigurable antenna structure conformal to a spherical surface with radius $R$. The locations of candidate radiating elements are parameterized in a spherical coordinate system [6], where each candidate position is specified by the azimuth angle $\phi$ and the elevation angle $\theta$. The spherical surface consists of electrically driven candidate ports or pixel-level radiating elements, which can be implemented using liquid-metal-based elements or RF-switch-based reconfigurable pixel antennas. Their common objective is to provide a high-resolution, electrically controllable, and rapidly switchable set of radiating elements as the physical basis for 3D spatial reconfiguration. By activating radiating elements in different regions on the spherical surface, the 3D SFAS can realize continuous reconfigurable region switching, thereby enabling array-level spatial reconfiguration. Furthermore, within the selected activated regions, the system can perform fine-grained element-level reconfiguration by adjusting the number of activated radiating elements, aperture size, effective array topology, and radiation characteristics. Therefore, unlike 2D FAS, which mainly rely on port or radiating-element position switching, the operating state of a spherical fluid antenna is jointly determined by the activated spherical region and the effective aperture configuration. This extends fluid antenna reconfiguration beyond 2D port switching toward the dynamic adjustment of the effective array location, orientation, aperture size, and radiation characteristics in 3D space.

To enable dynamic activation of the spherical fluid antenna, a backend digital control module is deployed inside the sphere and coordinated with an RF switch matrix and a reconfigurable feeding network. Based on channel state information (CSI), user spatial locations, and system task requirements, the controller generates array-level region selection and element-level state configuration commands. These commands are delivered to the spherical antenna surface to activate candidate radiating elements within the selected regions. This control procedure provides the hardware implementation basis for joint array-level and element-level reconfiguration.

### B. Array-Level Reconfiguration over the Spherical Aperture

For array-level reconfiguration, the 3D spherical fluid antenna electronically activates different regions on the spherical surface to dynamically reconfigure the effective array location and aperture distribution. Unlike conventional FPAs, whose aperture locations are predefined, and mechanically movable antennas, which rely on physical translation or rotation, the proposed architecture achieves spatial reconfiguration without mechanical movement. Specifically, it electronically activates radiating elements in selected regions on the spherical surface, thereby reconfiguring the effective aperture location and aperture distribution in 3D space. One or more regions on the spherical surface can be activated simultaneously, and these activated regions determine the effective aperture distribution and dominant coverage direction of the current array. When the propagation environment or service requirements change, the system can rapidly switch the activated regions on the spherical surface, thereby relocating the effective aperture and adapting the beam coverage. Thus, array-level reconfiguration is no longer limited to two-dimensional position switching, but is extended to effective aperture reconfiguration in 3D space.

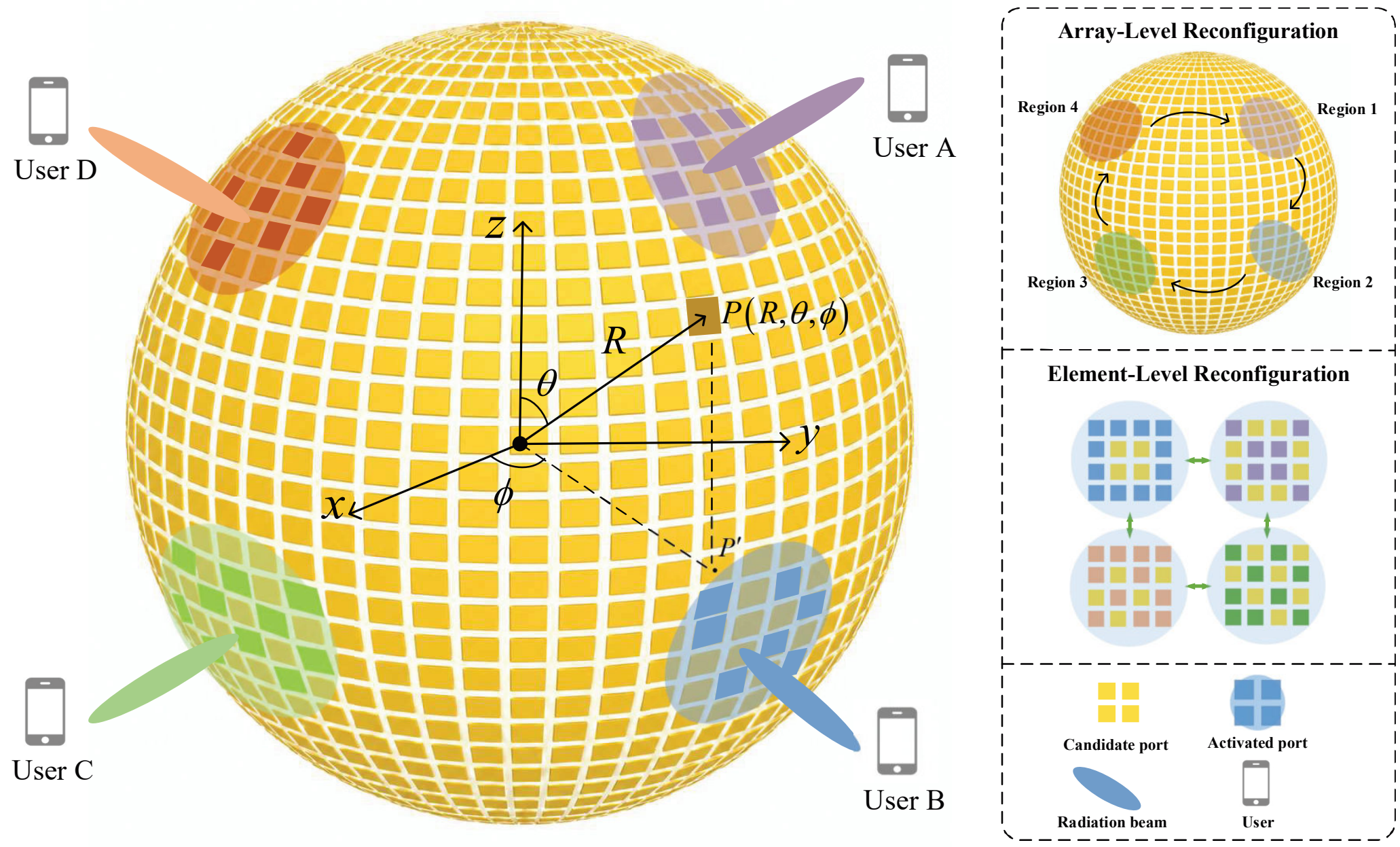

Fig. 1. The system model of 3D spherical fluid antenna.

### *C. Element-Level Reconfiguration over the Spherical Aperture*

In addition to array-level spatial reconfiguration, the 3D spherical fluid antenna supports fine-grained element-level reconfiguration within the activated regions, including adjustments to the geometric topology, radiation mode, polarization state, and local electromagnetic response. This capability enables further optimization of the beam shape, sidelobe distribution, polarization matching, and spatial coverage within the selected regions. In practice, element-level reconfiguration can be implemented using RF switches, reconfigurable feed networks, or port-selection mechanisms to selectively activate radiating elements at different positions, thereby dynamically modifying the local geometric topology and electromagnetic response. Overall, the proposed architecture establishes a joint array-level and element-level reconfiguration framework. Through array-level reconfiguration, the system provides large-scale spatial adaptation by selecting effective regions on the spherical surface, whereas through element-level reconfiguration, it refines the local electromagnetic response within the selected regions [7]. Together, these two reconfiguration levels constitute the core reconfigurability of the 3D spherical fluid antenna and provide unified hardware support for adaptive communications in complex 3D propagation environments.

### *D. Comparison with Existing Reconfigurable Antennas*

Table I compares several representative reconfigurable antenna architectures. Both 6DMA and rotatable antennas [8] primarily rely on mechanical translation or rotation to achieve spatial reconfiguration. Although these architectures provide additional spatial DoFs, their reliance on mechanical movement may limit the reconfiguration speed and lead to increased control complexity, deployment overhead, and reliability concerns. In contrast, 2D FAS can achieve faster reconfiguration through electronic port or position selection; however, their reconfigurable space is mainly confined to one- or two-dimensional apertures, which limits their adaptability to 3D propagation environments. The proposed 3D SFAS enables joint reconfiguration of the effective aperture location, array topology, and radiation characteristics through element-level or array-level activation on the spherical surface. Therefore, it provides a promising architecture for achieving faster reconfiguration response, higher reconfiguration accuracy, and enhanced 3D spatial reconfigurability in future reconfigurable wireless systems.

## III. Advantages and Performance Gains of 3D Spherical Fluid Antennas

### *A. Flexible Beamforming Enabled by Joint Array-Level and Element-Level Reconfiguration*

The 3D spherical fluid antenna enables enhanced flexible beamforming through joint array-level and element-level reconfiguration. The activated regions determine the effective aperture location and size of the current effective array, enabling adaptive adjustment of the beam direction, beam shape, and spatial coverage according to the propagation environment and service requirements. In conventional planar arrays, large-angle beam steering generally incurs scan loss due to reduced aperture projection and element-pattern degradation. In contrast, a 3D SFAS can activate the surface region oriented toward the target user, thereby dynamically relocating the effective aperture in the desired direction. This region-aware activation reduces the need for extreme-angle steering and helps maintain a high projected aperture gain over a broader 3D angular range. By coordinating element-level and array-level reconfiguration, the system can jointly optimize the effective aperture region, array topology, and local electromagnetic response. Such dual-scale reconfiguration enables highly flexible beamforming and more refined control of radiation characteristics.

TABLE I
COMPARISON OF DIFFERENT RECONFIGURABLE ANTENNAS

| Schemes | Reconfigurable Parameters | Reconfiguration Mechanism | Response Speed | Reconfiguration Accuracy | Control Complexity | 3D Spatial DoF |
|---|---|---|---|---|---|---|
| 6DMA | Position and orientation | Mechanical translation and rotation | Low to medium | Medium | High | High |
| Rotatable antenna | Array orientation | Mechanical rotation | Low | Low to medium | Medium | Medium |
| 2D FAS | 1D/2D port or position, radiation pattern | Fluidic or port selection | Medium to high | Medium | Medium | Limited |
| **3D SFAS** | **3D aperture position, topology, radiation** | **Spherical pixel-level antenna activation** | **High** | **High** | **Medium to high** | **High** |

Such flexible beamforming can rapidly respond to channel variations, mitigate beam misalignment, support continuous region selection and precise beam alignment across the spherical surface, and provide high spatial resolution for adjusting the effective aperture and beam characteristics. As a result, the received signal power, spectral efficiency, spatial coverage, interference suppression, and link robustness can be improved, enabling efficient adaptive communications in complex and dynamic 3D wireless environments.

### *B. Multi-Region Activation for Concurrent Multi-Beam Transmission in 3D Space*

By electronically activating multiple spatially separated regions on its spherical surface, a 3D SFAS can form several effective sub-apertures and thereby support concurrent multi-beam transmission. Each activated region can generate an independent beam toward a different spatial direction. As a result, the system is no longer limited to a single communication link or a single service direction, but can simultaneously cover multiple users, nodes, or targets distributed in 3D space. In multi-user scenarios, the system can select multiple favorable activation regions on the spherical surface according to the spatial distribution of users, thereby forming beams directed toward the corresponding users. Since the spherical structure can cover the full 3D angular domain, users located at different azimuth and elevation angles can be concurrently served through multiple effective sub-apertures. This capability helps improve spatial multiplexing and enhance system throughput in dense-access scenarios.

Moreover, multi-region activation is beneficial for multidirectional multi-link communication and multi-functional array operation. In space-air-ground integrated networks, communication nodes need to establish simultaneous links with satellites, UAVs, terrestrial base stations, vehicles, and ground users. The 3D SFAS can simultaneously activate regions oriented toward different spatial directions, thereby supporting concurrent multi-link access, access-backhaul coordination, and flexible spatial resource scheduling in complex 3D network environments. Meanwhile, different activated regions can be assigned to different functions, such as high-rate data transmission, target sensing, localization, environmental probing, and interference monitoring. Depending on task-specific requirements, each activated region can be independently configured in terms of its local geometric topology, beam shape, and radiation characteristics, thereby providing a flexible hardware foundation for integrated sensing and communication and environment-aware interaction with the wireless environment.

### *C. Blockage-Adaptive Spatial Channel Reconfiguration*

For conventional fixed-position arrays, both the aperture geometry and spatial location are fixed after deployment. When the dominant propagation path is blocked, the system can typically adjust only the beamforming weights over the fixed aperture or switch among predefined beams, offering limited capability for spatial channel reconfiguration. Although 2D FAS can perform port or position switching on planar apertures, their candidate locations remain confined to one- or two-dimensional structures, which limit their ability to provide sufficiently diverse alternative propagation paths in complex 3D blockage scenarios.

In contrast, the 3D SFAS can dynamically reconfigure the effective aperture across the spherical surface, thereby providing a more flexible hardware architecture and transmission scheme for adaptive spatial channel reconfiguration. When the propagation path associated with a given activated region is blocked, the system can switch to another candidate region on the spherical surface based on channel state information or link-quality assessment, thereby forming a new effective aperture and establishing an alternative propagation path. Meanwhile, element-level reconfiguration can further adjust the local topology and radiation characteristics, thereby improving effective path gain. Therefore, the proposed architecture extends channel adaptation beyond beamforming-weight optimization on a fixed aperture to the joint reconfiguration of the effective aperture location, local electromagnetic response, and spatial propagation paths, thereby improving link reliability under blockage conditions.

### D. Flexible Effective Array-Aperture Reconfiguration

The 3D spherical fluid antenna enables flexible spatially reconfigurable communications through adaptive effective array-aperture configuration, including the number of activated radiating elements, aperture size, array topology, and radiation characteristics. Based on user distribution, channel conditions, blockage states, and communication tasks, the system can dynamically optimize its effective array-aperture configuration to improve spatial adaptation. The 3D SFAS supports both switching among predefined uniform array topologies and dynamic optimization of nonuniform array topologies. For example, annular activation can form a topology analogous to an effective uniform circular array; activation along meridians, parallels, or arcs can form equivalent linear or curved arrays; and contiguous local activation can construct planar or conformal aperture configurations. By jointly optimizing the selected element positions and beamforming vectors, the system can form adaptive nonuniform array topologies. Moreover, the 3D spherical fluid antenna can simultaneously activate multiple spatially separated regions on the spherical surface to form distributed subarray configurations, thereby supporting spatial multiplexing and concurrent multi-beam transmission. The number of activated elements and the effective aperture size can also be flexibly configured based on service requirements and resource allocation among different aperture regions. The 3D spherical fluid antenna provides enhanced spatial flexibility by selecting effective apertures over both azimuth and elevation dimensions. By orienting different array-aperture configurations toward arbitrary user or target directions, it can maintain favorable aperture projection across a broader spatial domain, thereby improving reconfigurability and robustness in complex environments.

### E. High-Resolution 3D Spatial Reconfiguration

As future wireless systems evolve toward millimeter-wave, terahertz, and extremely large-scale MIMO regimes, wireless beams become increasingly narrow, necessitating more precise control over beam direction, aperture placement, and energy distribution. The 3D spherical fluid antenna can achieve high-resolution spatial reconfiguration through densely deployed candidate ports or pixel-level radiating elements on the spherical surface. By selecting regions across both azimuth and elevation dimensions, the system can continuously adjust the position, size, and topology of the effective aperture. This capability improves beam alignment accuracy, spatial resolution, received signal power, and spectral efficiency, while supporting smoother beam tracking and faster link adaptation in dynamic scenarios. The 3D spherical fluid antenna also exhibits considerable potential for near-field communications. As the carrier frequency and array aperture increase, users and targets are more likely to reside within the near-field region, where the channel depends on the angular domain, propagation distance, and wavefront curvature. Through spherical region selection and aperture reconfiguration, the system can adjust the position, orientation, shape, and energy distribution of the effective aperture, thereby improving spatial matching with near-field users. This capability helps distinguish users with similar angular directions but different distances, focus energy toward target regions, reduce energy leakage, and enhance localization, range estimation, and 3D sensing for ISAC.

## IV. Applications of 3D Spherical Fluid Antennas

Owing to these architectural advantages, 3D spherical fluid antennas are expected to show great potential in dynamic, multi-node, multi-directional, and multi-functional wireless scenarios. As illustrated in Fig. 2, representative applications include space-air-ground integrated networks, vehicular and high-mobility communications, integrated sensing and communication, and emergency communications.

### A. Space-Air-Ground Integrated Networks

In space-air-ground integrated networks, communication nodes are distributed across multiple spatial layers, including satellites, high-altitude platforms, UAVs, terrestrial base stations, vehicles, and ground users, leading to highly dynamic 3D links. The 3D SFAS can rapidly activate candidate regions oriented toward different target nodes, thereby forming effective apertures that provide adaptive beam coverage for satellite, aerial, terrestrial, and lateral links. With multi-region activation, the 3D SFAS can further support concurrent multi-link access, access-backhaul coordination, and flexible spatial resource scheduling, thereby improving link reliability and spatial resource utilization in multi-layer networks.

### B. Vehicular and High-Mobility Communications

In vehicular and high-mobility communications, rapid vehicle movement, building-induced blockage, and dynamically changing road environments may result in frequent link interruptions. By deploying 3D SFAS on vehicles or roadside units, the system can rapidly adjust activated regions on the spherical surface to maintain adaptive beam coverage for mobile users and infrastructure nodes. When the dominant propagation path is blocked, the system can activate another region on the spherical surface to exploit alternative propagation paths, including reflected or scattered components. This capability improves link robustness and service continuity in high-mobility scenarios.

### C. Integrated Sensing and Communication

Integrated sensing and communication requires flexible support for data transmission, target detection, localization, and environmental sensing. Through multi-region activation, the 3D SFAS can allocate different spherical regions to distinct communication and sensing tasks, such as assigning one region to user data transmission and another to target detection or environmental scanning. The continuous candidate activation regions on the spherical surface enable high-resolution multi-directional sensing, thereby enhancing sensing accuracy and spatial coverage. Each activated region can be independently configured according to task-specific requirements, including its geometric topology, beam shape, and radiation characteristics. The 3D SFAS provides a spatially reconfigurable hardware foundation for environment-aware wireless communication systems, supporting flexible communication-sensing coordination while enhancing coverage and sensing resolution.

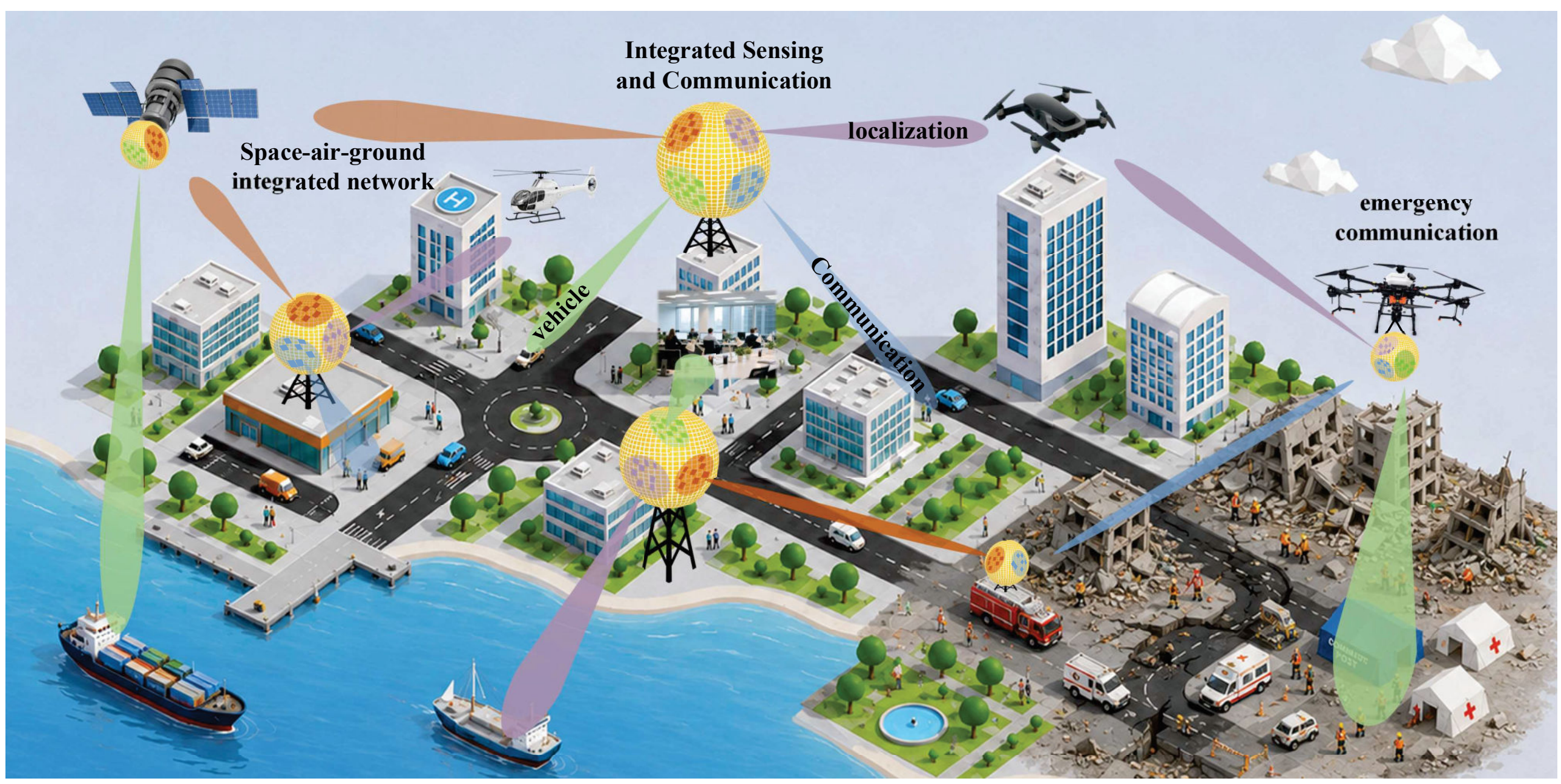

Fig. 2. Representative applications of 3D spherical fluid antennas.

### *D. Emergency Communications*

Emergency communication systems often operate under damaged or unavailable infrastructure, complex propagation environments, unstable link conditions, and rapidly changing node distributions. In scenarios such as earthquakes, floods, fires, landslides, or building collapses in urban areas, existing terrestrial base stations may become unavailable, while rescue vehicles, UAVs, temporary base stations, satellite links, and ground terminals must rapidly establish reliable connections in complex 3D environments [9]. Therefore, emergency communication systems require not only rapid deployment, but also flexible link reconfiguration to adapt to blockage, node mobility, and environmental changes. The multi-region activation capability of 3D SFAS enables simultaneous support for multi-directional communication links. For example, one region on the spherical surface can be assigned to UAV or satellite backhaul connectivity, another can be configured to serve ground rescue personnel or mobile terminals, and additional regions can support sensing, localization, or environmental probing. When the dominant propagation path is blocked by buildings, terrain, or other obstacles, the system can rapidly activate another region on the spherical surface to exploit alternative propagation paths, including reflected or scattered components, thereby improving link robustness and service continuity. Meanwhile, element-level reconfiguration can adaptively configure the number of activated elements, effective aperture size, array topology, and radiation characteristics based on task-specific requirements. This allows the system to support different communication, sensing, and localization tasks without replacing the antenna hardware. By supporting multi-region, multi-link, and multi-functional coordination, 3D SFAS can provide a flexible reconfigurable hardware foundation for emergency communication networks.

## V. Challenges and Open Problems

Despite the promising potential of 3D spherical fluid antennas, several open issues remain to be further investigated. First, low-overhead CSI acquisition is required to characterize the coupling among activated spherical regions, element-level reconfiguration states, and complex 3D propagation environments. Given the large number of candidate regions and reconfigurable states on the spherical surface, efficient beam and region management strategies are essential for reducing training overhead and enabling real-time reconfiguration. Second, efficient joint control and resource allocation strategies are needed to coordinate array-level region selection and element-level state reconfiguration based on user locations, CSI, and service requirements. Finally, practical deployment requires the co-design of the spherical antenna structure, RF switch network, reconfigurable feed architecture, and multi-link signal integration and processing mechanisms for array-level and element-level reconfiguration.

## VI. Performance Evaluation

To evaluate the proposed 3D SFAS, we compare spectral efficiency and normalized directional gain under the same transmit power and number of activated radiating elements. The FPA uses a uniform planar aperture, the 2D FAS selects candidate ports on a planar surface, and the 3D SFAS selects candidate regions on a spherical surface. For joint reconfiguration, both the activated spherical region and the element subset within it are optimized.

As shown in Fig. 3, the spectral efficiency of all considered schemes increases with SNR. The fixed planar array exhibits the lowest spectral efficiency, since it lacks physical aperture reconfiguration capability. The 2D FAS scheme achieves a moderate performance gain through port or position selection, but its candidate positions are restricted to a two-dimensional aperture, which limits its adaptability to 3D propagation environments. In contrast, through array-level reconfiguration, the 3D SFAS can select a more favorable effective-aperture region on the spherical surface, thereby achieving higher channel gain. Furthermore, joint array-level and element-level reconfiguration optimizes the effective-aperture location and local electromagnetic response, leading to the highest spectral efficiency across the entire SNR range.

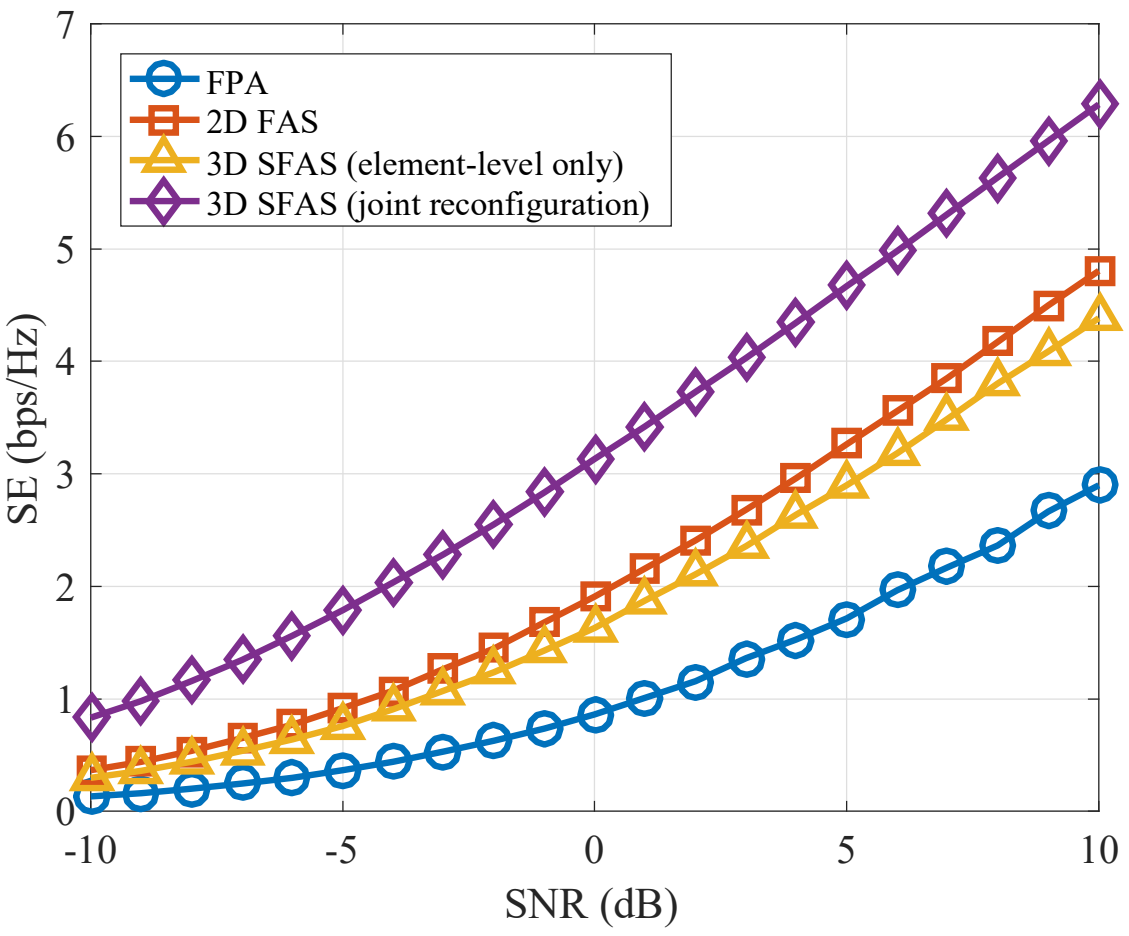

Fig. 3. Spectral efficiency versus SNR.

Figure 4 compares the normalized directional gain of different architectures in a two-user scenario, where User A is located at an angle of $-\pi/4$ and User B is located at an angle of $\pi/3$. The fixed planar array and the 2D FAS scheme exhibit limited gain toward User B, mainly due to aperture-projection loss and restricted candidate-position selection on a two-dimensional aperture, respectively. In contrast, the 3D SFAS can activate a region oriented toward the user located at a large angular offset, thereby achieving higher directional gain. Furthermore, by activating two regions simultaneously, the proposed architecture can form beams directed toward User A and User B, demonstrating its concurrent multi-region beamforming capability. Since equal power allocation is adopted in the two-region scheme, each individual beam has a lower peak gain than that in the single-region scheme dedicated to a single user. Nevertheless, this two-region activation scheme enables concurrent multi-user coverage and highlights the flexibility of SFAS for multi-directional communications.

## VII. Conclusions

This article discussed a 3D SFAS for future flexible and spatially reconfigurable wireless communication systems. By electronically activating different regions on the spherical surface and configuring the element-level states within the selected regions, the proposed architecture enables joint array-level and element-level reconfiguration without relying on mechanical movement. This dual-scale reconfiguration can enable highly flexible beamforming, multi-region-enabled multi-beam transmission, blockage-adaptive reconfiguration, flexible effective-aperture reconfiguration, and high-resolution 3D spatial control. These capabilities highlight the potential of 3D SFAS in space-air-ground integrated networks, vehicular and high-mobility communications, and ISAC systems. Numerical results demonstrate the potential benefits of 3D spherical fluid antennas for future spatially reconfigurable wireless communication systems. Overall, 3D SFAS provide a promising pathway for extending FAS design beyond 2D position selection toward comprehensive 3D spatial reconfigurability.

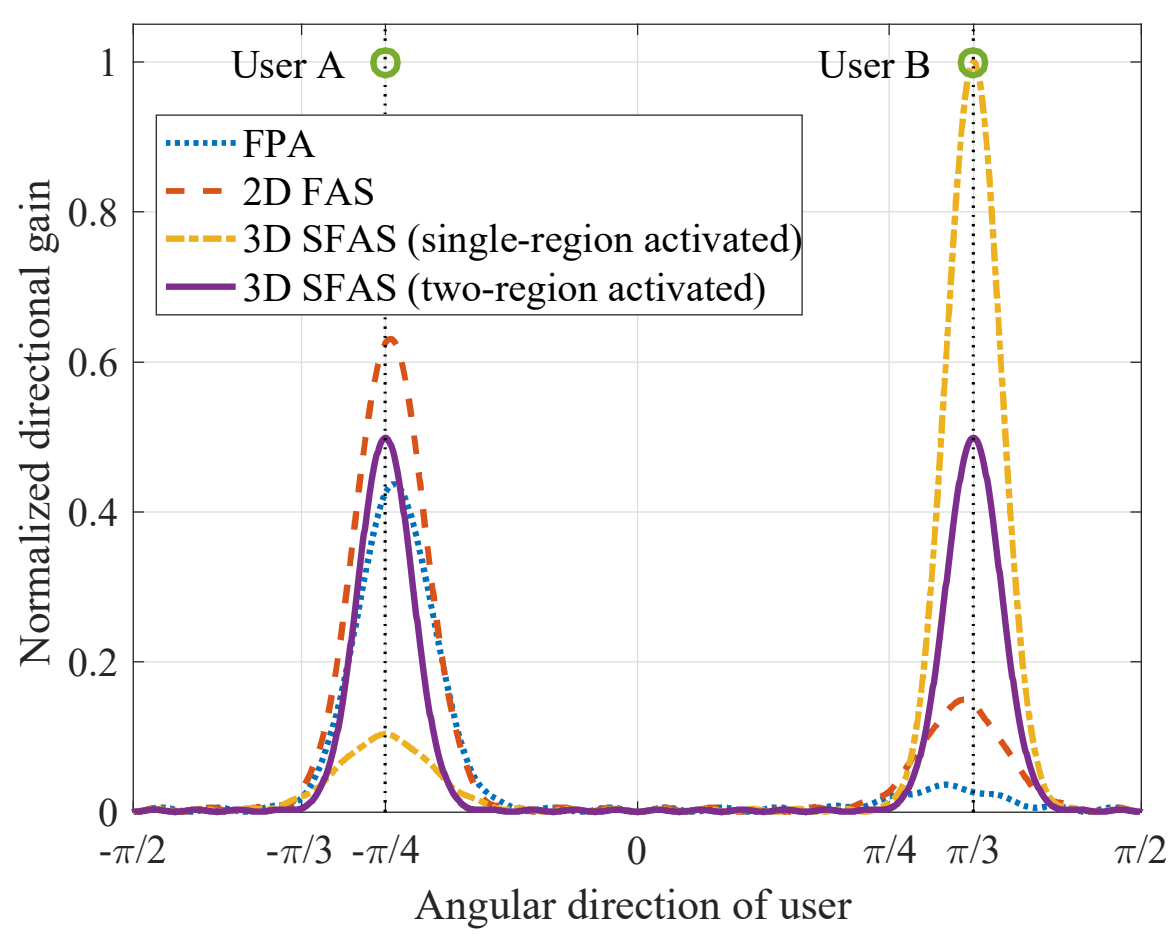

Fig. 4. Multi-region beamforming capability of 3D SFAS.